\documentclass[aps,twocolumn,amsfonts]{revtex4}
\begin{document}
\newcommand{\bb}{\begin{equation}}
\newcommand{\ee}{\end{equation}}
\newcommand{\eqb}{\begin{eqnarray}}
\newcommand{\eqf}{\end{eqnarray}}

\preprint{}
\title{U(1) Noncommutative Gauge Fields and Magnetogenesis}
\author{J.  Gamboa}
\email{jgamboa@lauca.usach.cl}
\affiliation{Departamento de F\'{\i}sica, Universidad de Santiago de Chile,
Casilla 307, Santiago 2, Chile}
\author{Justo  L\'opez-Sarri\'on}
\email{justo@dftuz.unizar.es}
\affiliation{Departamento de F\'{\i}sica, Universidad de Santiago de Chile,
Casilla 307, Santiago 2, Chile} 
\begin{abstract}
 The connection between the Lorentz invariance violation in the
 lagrangean context and the quantum theory of noncommutative fields is
 established for the $U(1)$ gauge field. The modified Maxwell
 equations coincide with other derivations obtained using different
 procedures. These modified equations are interpreted as describing
 macroscopic ones in a polarized and magnetized medium. A tiny
 magnetic field (seed) emerges as particular static solution  that
 gradually increases once the modified Maxwell equations are solved as
 a self-consistent equations system. 
\end{abstract}
\pacs{PACS numbers:}
\maketitle
\section{Introduction}

In the last few years several authors have suggested possible Lorentz invariance violation at  quantum field theory and particle physics level \cite{review}. If this theoretical suggestions are true, then our conception on Lorentz invariance and spacetime would be only approximate ideas coming from a more fundamental -- still unknown-- structure. From this point of view, these results could be another indication that the present relativistic quantum field theories descriptions would correspond to effective field theories.

Two important approach going beyond to the standard Lorentz symmetry are doubly special relativity \cite{amelino} and the extended standard model \cite{alan} which are  proposals which try to give an answer to largely unsolved problems in high energy physics such as, ultra high cosmic rays \cite{review2}, matter-antimatter asymmetry 
\cite{dine}, primordial magnetic field \cite{rubi,dolgov, ralston,berto,sheik}.  

A third possibility is quantum theory with noncommutative fields which
has been proposed in 
\cite{nos}, where the Lorentz symmetry is broken by modifying the
canonical commutators including an ultraviolet and infrared scales. As
we will consider an expanding universe surrounded by radiation,
one could guess that a Bohr-Oppenheimer approach \cite{john1} naturally should
generate a geometrical connection which produces a non-commutativity
in the momenta space at the quantum level.

The goals of the present paper are two; firstly, we will investigate
the connection between the Kosteleky {\it et. al.} approach to quantum
field theory and quantum theory with noncommutative fields for the
particular context of the abelian gauge field and secondly, once the
equivalence between both approaches is proven, we will explain some
consecuences for the primordial magnetic field. 

 More precisely, we will show  that the modified Maxwell equations
--that are the same
equations found by Carroll {\it et. al.}-- contain as a solution a
universe  filled with a tiny magnetic field. However, once this
tiny magnetic field is given, the modified Maxwell equations generate {\it per se}
a  very natural self-interacting mechanism which is an alternative to the dynamo 
mechanism.

The paper is organized as follows: in section 2 we will prove the equivalence between the Kosteleck\'y {\it et al} approach and quantum theory with noncommutative fields for an abelian gauge field. In section 3, we will reinterpret the modified Maxwell equations as macroscopic ones which suggests, in section 4, the way in which a primordial magnetic field might appear. Finally in section 5, the conclusions and other possible physical implications are given. 

\section{$U(1)$ Gauge Field as a Noncommutative Gauge Field }

In order to discuss the $U(1)$ gauge field as a noncommutative one, let us recall the hamiltonian formulation of the abelian gauge field. 

The lagrangian for an abelian gauge field
\begin{equation}
{\cal L} = -\frac{1}{4} F_{\mu \nu}F^{\mu \nu}, \label{1}
\end{equation}
is invariant under the gauge transformation 
\begin{equation}
A_\mu \rightarrow A_\mu + \partial_\mu \Lambda. \label{2}
\end{equation}

Thus, (\ref{1}) has two symmetries, namely,  the gauge and Lorentz symmetry. 

The hamiltonian analysis yields to the canonical momentum
\begin{equation}
\pi^\mu = F^{0\mu},  \label{3}
\end{equation}
and, therefore, one has the primary constraint 
\begin{equation}
\pi^0 =0. \label{4}
\end{equation}

Thus, the canonical hamiltonian is 
\begin{equation}
H =  \int d^3x \left( \frac{1}{2}{\vec \pi}^2 + \frac{1}{2}{\vec B}^2+A_0\,\vec\nabla\cdot\vec\pi\right), \label{5}
\end{equation}
and the preservation of (\ref{3}) implies 
\begin{equation}
{\dot \pi}^0 (x) = \left[ \pi^0 (x), H\right] = \nabla . ~{\vec \pi}, \label{6}
\end{equation}
{\it i.e.} the secondary constraint is the Gauss' law.

The Gauss' law is a first order constraint and, from the hamiltonian point of view, it 
generates the gauge symmetry (\ref{2}). 

Once the constraints are found, the gauge field is quantized changing the Poisson brackets 
\begin{eqnarray}
\left[ A_i ({\vec x}),A_j ({\vec y})\right]_{PB} &=&0, \nonumber
\\
\left[ A_i ({\vec x}),\pi^j ({\vec y})\right]_{PB} &=& \delta^j_i \delta ({\vec x}-{\vec y}), \label{7}
\\
\left[ \pi_i ({\vec x}),\pi_j ({\vec y})\right]_{PB}&=&0, \nonumber
\end{eqnarray}
by commutators according to the rule $[,]_{PB} \rightarrow [,]/i\hbar$.

The $U(1)$ noncommutative field is constructed by deforming the previous Poisson algebra as follows, 
\eqb
\left[ A_i ({\vec x}),A_j ({\vec y})\right]_{PB} &=&0, \nonumber
\\
\left[ A_i ({\vec x}),\pi_j ({\vec y})\right]_{PB} &=& \delta_{ij} \delta ({\vec x}-{\vec y}), \label{8}
\\
\left[ \pi_i ({\vec x}),\pi_j ({\vec y})\right]_{PB}&=&\theta_{ij}\delta ({\vec x}-{\vec y}), \nonumber
\eqf
where $\theta$ is the most general antisymmetric three dimensional matrix. 

Although the Poisson brackets (\ref{8}), of course, break Lorentz invariance, one can retain the gauge symmetry. Indeed, in order to do that, we must modify the 
Gauss' law appropriately. 

Thus, the modified Gauss' law  should be 
\bb
\chi = \partial_i \pi_i + \text{something}, \label{9}
\ee
where \lq \lq something" represents the modified term, which is constrained to
satisfy the relations,
\begin{eqnarray}
\delta A_i(\vec x)=\left[A_i(\vec x), \Delta_\alpha\right]_{PB} &=& 
\partial_i\alpha(\vec x), \label{12}
\\
\delta \pi_i(\vec x)=\left[\pi_i(\vec x), \Delta_{\alpha}\right]_{PB}&=&0,\label{13}
\end{eqnarray}
where $\Delta_\alpha$ is defined as 
\bb
\Delta_{\alpha}\equiv -\int d^3x\, \alpha(x)\chi(x),\label{13a}
\ee
and where $\alpha(x)$ is an arbitrary real function.

Now, it is easy to see that the modified Gauss' law must be given by the constraint,
\bb
\chi= \nabla . ~{\vec \pi} - {\vec \theta}.~{\vec B}, \label{11}
\ee
where $\theta_{ij} A_j= \epsilon_{ijk}\theta_k A_j= -{\vec \theta}\times {\vec A}$,
and therefore the gauge transform operator (\ref{13a}) can be written as,
\eqb
\Delta_{\alpha}&=& -\int d^3x\, \alpha(\vec x)\left\{\vec\nabla\cdot\vec\pi
-\vec\theta\cdot\vec B\right\} \nonumber
\\
&=&\int d^3x\,\alpha(\vec x)\vec\nabla\cdot\left(\vec \pi
+ \vec\theta\times\vec A\right),\label{14}
\eqf
and the modified total hamiltonian which generalizes the $U(1)$ system should be, 
\bb
H= \int d^3x\, \left(\frac{1}{2} {\vec \pi}^2 + \frac{1}{2}{\vec B}^2 + A_0 \nabla \left( {\vec\pi} + 
{\vec \theta}.\times {\vec A}\right) \right). \label{15}
\ee

Using (\ref{15}) one finds that the equations of motion are
\eqb
{\dot A}_i &=& \left[ A_i, H\right]_{PB} = \pi_i - \partial_i A_0, \label{16}
\\
{\dot \pi}_i &=& \left[ \pi_i, H\right]_{PB} = 
({\vec\pi}\times {\vec \theta})_i - (\nabla \times {\vec B})_i. \label{17}
\eqf

The first equation, of course, is, basically, the standard definition of electric field, {\it i.e.},
\bb
\pi_i = - E_i \equiv \dot A_i + \partial_i A_0,\label{17a}
\ee
and, hence, the second one
\bb
\frac{\partial {\vec E}}{\partial t} = \nabla \times {\vec B} + {\vec E} \times {\vec \theta}, \nonumber  
\ee 
is the modified Ampere's law. 

The remaining equations, namely
\eqb
\nabla .~{\vec B}&=& 0, \label{18}
\\
\nabla \times {\vec E}&=& - \frac{\partial {\vec B}}{\partial t}, \label{19}
\eqf
have no changes. And as we said above, the Gauss' law is written as,
\bb
\nabla.~\vec E +\vec\theta.~\vec B = 0. \label{18a}
\ee

Now, let us find out the lagrangian where these equations come
from. In order to do that, we should find a set of canonical
conjugated variables to $A_i$. This is, in fact, easy to find by taking in account (\ref{8}). 

We find that these new variables are just,
\bb
\tilde{\pi}_i\equiv \pi_i + \frac{1}{2}(\vec\theta\times \vec A).\label{18b}
\ee

From these results one gets the lagrangian as follows; firstly, we write
\eqb
L &=&  \int d^3 x ~\tilde{\pi}_i {\dot  A}_i - H \nonumber
\\
&=& \int d^3x \left( {\vec E} - \frac{1}{2} \vec\theta \times {\vec A}\right) \left({\vec E} + 
\vec\nabla A_0 \right) -H \nonumber
\\
&=& \int d^3x \left( {\vec E}^2 - {\vec B}^2 + \frac{1}{2}A_0 {\vec \theta}.~{\vec B} 
- \frac{1}{2}{\vec A}. ~{\vec \theta} \times {\vec E}) \right). \nonumber 
\\ 
\label{20}
\eqf

Using the standard definition for $F_{\mu \nu}$ and ${\tilde F}^{\mu \nu}\equiv
\frac{1}{2}\epsilon^{\mu\nu\lambda\rho}F_{\lambda\rho}$, one finds that the 
lagrangean is 
\bb
L = \int \left( -\frac{1}{4} F_{\mu \nu} F^{\mu \nu} + \frac{1}{2} \theta_\mu 
{\tilde F}^{\mu \nu}A_\nu\right), \label{21}
\ee 
where in our case the four-vector $\theta_\mu$ is $(0, {\vec \theta})$. 

The modified Maxwell equations obtained in this paper were derived
from a completely different point of view to the used in \cite{carroll}.  Our calculation shows explicitly the connection between these two  apparently non-related approaches.

A discussion on physical aspects related to the propagation of the light for these modified photons and other systems can be found in \cite{sasha}. In particular, the dispersion relation in this space-like approach is,
\begin{equation}
\omega^2_{\pm}= \vec k^2 +\frac{1}{2}\vec\theta^2\pm \sqrt{(\vec
k\cdot\vec\theta)^2 +\frac{1}{4}(\vec\theta^2)^2}. \label{22}
\end{equation}

\section{Interpreting the Modified Maxwell Equations}

In this section we will give a physical interpretation of the modified Maxwell equations. 

Let us start assuming that possible Lorentz invariance violation
processes could have occurred  in the early universe and some tiny
relics could be observable presently. As photons are the most abundant
particles in the present universe, one can think of that some relics could be accessible via electromagnetic processes. 

It is interesting to note that the Modified Maxwell equations contain a 
\lq \lq source" term $-{\vec \theta}. ~{\vec B}$ and ${\vec
  \theta}\times {\vec E}$ that can be interpreted as polarization charges and induced currents on a medium in a similar way to the standard electromagnetic theory. 

Therefore, these modifications of the Maxwell equations suggest us to
consider a sort of modified displacement vector (${\vec D}$) and magnetic field vector (${\vec H}$) where 
\eqb
{\vec D} &=& {\vec E} - {\vec \theta}\times {\vec A}, \label{25}
\\
{\vec H} &=& {\vec B} + {\vec \theta}A_0. \label{26}
\eqf

Using these definitions the modified Maxwell equations can be written as the standard Macroscopic Maxwell equations in a medium, {\it i.e.}
\eqb
\nabla . ~{\vec D} &=& \rho, \nonumber 
\\
\nabla \times {\vec H} &=& \frac{\partial {\vec D}}{\partial t} + {\vec J}, \nonumber 
\\
\nabla \times {\vec E} &=& -\frac{\partial {\vec B}}{\partial t}, \label{27}
\\
\nabla .~{\vec B} &=& 0, \nonumber
\eqf
where $\rho$ and ${\vec J}$ are possible external sources. 

One should note that the polarization and magnetization vectors
\eqb
{\vec P} &=& - {\vec \theta}\times {\vec A}, \label{28}
\\
{\vec M} &=&  {\vec \theta}A_0, \label{29}
\eqf
are not gauge invariants, however this is not important because the
physically relevant quantities are $\nabla .~{\vec P}$ and ~~$\nabla
\times {\vec M}-\dot{\vec P}$ which are, in fact, gauge invariants \cite{jackson}.  

It is interesting to note that in the static scenario, the electrostatic and magnetostatic effects appear mixed and, therefore, the presence of  polarization  implies a magnetization of a medium and vice versa.

This result is a consequence of the modified Maxwell theory and it is not true in the conventional electromagnetic theory.

\section{Origin of the Primordial Magnetic Field}

The structure of the above  modified Maxwell equations might give a
guess on the origin of the intergalactic magnetic field as well as, it
might provide of a simple alternative argument to dynamo mechanisms discussed in the literature (see {\it e.g.} \cite{rubi,dolgov,ralston}). 

In order to explain this fact, let us suppose that it is generated a seed magnetic field ($\vec B^{(0)}$) parallel to $\vec\theta$. If this processes take place during a long time, we can suppose that only the stationary equations are the important ones, {\it i.e.},
\begin{eqnarray}
\vec\nabla\cdot\vec E +\vec\theta\cdot\vec B & = & 0\label{b1}\\
\vec\nabla\times\vec B + \vec\theta\times \vec E & = & 0\label{b2}\\
\vec\nabla\times \vec E &=&0\label{b3}\\
\vec\nabla\cdot\vec B &=&0\label{b4}
\end{eqnarray}
Then, a constant magnetic field is a solution of all equations at
zeroth order in $\theta$ (or in the spatial scale $r$). However, if we consider the first order in $\theta$ (or we displace a distant $r$ from the origin) the equation (\ref{b1})
will demand an electric field,
\begin{equation}
\vec E^{(1)} = -\frac{1}{3} B^{(0)}(\theta r)\hat r
\end{equation}
The next order in $\theta$ is given by the equation (\ref{b1}) (of course, in this
procces equations (\ref{b3}) and (\ref{b4}) are always present). 

This equation 
generates a second order correction in the magnetic field,
\begin{equation}
\vec B^{(2)} = -\frac{1}{30} B^{(0)} (\theta r)^2 (7\cos\tilde\theta\,\hat e_{r} 
- 9 \sin\tilde\theta \,\hat e_{\tilde\theta})
\end{equation}
where $(r,\tilde\theta,\tilde\phi)$ and $(\hat e_r,\hat e_{\tilde\theta},\hat 
e_{\tilde\phi})$ are the spherical coordenates and their unit vector fields. We can 
follow this expansion in order to get all orders in $\theta$, and one should obtain
a potential series for $\vec B$ and $\vec E$, {\it i.e.},
\begin{eqnarray}
\vec B &=& \vec B^{(0)} + \vec B^{(2)} + \vec B^{(4)} +\dots+\vec B^{(2n})+\dots\\
\vec E &=& \vec E^{(1)} + \vec E^{(3)} + \vec E^{(5)} +\dots+\vec E^{(2n+1})+\dots\\
\end{eqnarray}
where the superindices stand for the order in $\theta$. It should be noted that $E\sim
\theta B$, which means that the electric field is always a lower order
of magintude
that the magnetic field, according to the experimental fact. 

The possibility for these expansions to be divergent series suggests us that
the system might evolve to a stable state with permanent magnetic and/or electric
fields, in the similar way to ferromagnetic media. Then this mechanism would be a
possible candidate for an alternative explanation to the dynamo
mechanism of the primordial magnetic field observed in the universe.

\section{Conclusions and Other Physical Implications}

In this paper we have shown that deformation of the canonical
commutators for the electromagnetic field yields to a modification of
the Maxwell electrodynamics where electrostatics and magnetostatics
appears mixed \footnote{This fact has been noted recently in \cite{alan1}.}, this is direct consequence of the Lorentz invariance violation. 

Although the modified Maxwell equations are formally the standard
macroscopic ones, the underlying physics is quite different to the
conventional interpretation. Indeed, the mix between electrostatics
and magnetostatics induces as a consequence polarizations and
magnetizations and hence, 
physical electrical or magnetic fields. 

From the physical point of view, this is a very interesting new effect
because it could be the arena for the elusive primordial magnetic
field.  Indeed, as the universe expansion has spherical symmetry and
as the universe is made mainly of photons, then one can see the universe as a sort of  magnetized sphere.  If we assume that the electromagnetic fields are --as a first approximation-- static and the radius of the present universe is $a$, our universe should be filled with  a magnetic field like 
\bb
{\vec H} = -\frac{2}{3} {\vec M}, \label{31}
\ee
however, as ${\vec M}$ is proportional to $|{\vec \theta}|$ it is a very tiny energy scale -- like $\sqrt{\Lambda}$-- then also ${\vec H}$ should be a tiny magnetic field filling our present universe. 

In this sense, the modified electrodynamics --as a consequence of a tiny  Lorentz invariance violation--  might be a mechanism for the origin of the elusive seed field observed in galaxies. 

Possibles implications with the Born-Oppenheimer approximations \cite{john1} and other  nonperturbatives phenomena, will be discussed elsewhere. 

This work has been partially supported by the grants 1010596  from Fondecyt-Chile and MECESUP-USA-0108. We would like to thank A. A. Andrianov, F. M\'endez, 
J.P. Ralston and J. L. Cort\'es for discussions and comments.

\end{document}